\documentclass[12pt,a4paper]{article}
\usepackage{amsfonts,latexsym}
\usepackage{graphicx,color}
\usepackage{dcolumn}
\usepackage{graphicx}
\usepackage{amsmath}
\usepackage{amsfonts}
\usepackage{amssymb}

\renewcommand{\thefootnote}{\fnsymbol{footnote}}

\begin{document}

\vspace{12mm}

\begin{center}
{{{\Large {\bf Correlated stability conjecture for AdS black holes \\ in higher dimensional  Ricci cubic gravity}}}}\\[10mm]

{Yun Soo Myung$^a$\footnote{e-mail address: ysmyung@inje.ac.kr} and De-cheng Zou$^{b}$\footnote{e-mail address: dczou@yzu.edu.cn}}\\[8mm]

{${}^a$Institute of Basic Sciences and Department  of Computer Simulation, Inje University, Gimhae 50834, Korea\\[0pt] }

{${}^b$College of Physics and Communication Electronics, Jiangxi Normal University, Nanchang 330022, China\\[0pt]}
\end{center}
\vspace{2mm}

\begin{abstract}
We investigate  the correlated stability conjecture  for AdS black
holes obtained from the higher dimensional Ricci cubic  gravity.
It shows  that the
Ricci tensor perturbations exhibit unstable modes
for small AdS black holes when solving  Lichnerowicz equation, leading to Gregory-Laflamme instability. On the other hand, we find that a small black hole is thermodynamically unstable by showing  the negative heat capacity.  This suggests that the correlated stability conjecture holds for  AdS black holes in Ricci cubic gravity. Furthermore, we  find a newly non-AdS black hole
by solving  static  Lichnerowicz equations, confirming the threshold mass for Gregory-Laflamme instability.

\end{abstract}
\vspace{5mm}

\vspace{1.5cm}

\hspace{11.5cm}
\newpage
\renewcommand{\thefootnote}{\arabic{footnote}}
\setcounter{footnote}{0}


\section{Introduction}

 GR (general relativity) black holes could be justified  through  classical stability analysis by solving the linearized Ricci tensor equation $\delta R_{\mu\nu}(h)=0$. If one adopts the Regge-Wheeler gauge, two physical degrees of freedom for a massless spin-2 mode
could be captured by splitting  odd and even parities.
If a black hole is stable against the metric perturbation $h_{\mu\nu}$, it could  survive  as a physical black hole. It turned out that Schwarzschild black hole is stable against metric perturbations~\cite{Regge:1957td,Zerilli:1970se,Vishveshwara:1970cc}, even though it is thermodynamically unstable because its heat capacity is always negative. Rotating black hole solution of astrophysical relevance  is classically  stable~\cite{Teukolsky:1972my,Whiting:1988vc}, but its heat capacity is negative for small rotation $a$ and positive for large rotation $a$~\cite{Monteiro:2009tc}.
Investigating the stability analysis of the Schwarzschild-AdS (SAdS) black hole in Einstein gravity with a cosmological constant, one uses the linearized Einstein equation $\delta G_{\mu\nu}(h)=0$ to show its stability~\cite{Cardoso:2001bb,Ishibashi:2003ap}. This black hole is split thermodynamically  into small (unstable) and large (stable) black hole. Therefore, it was shown that the classical stability of these black holes is not related to with thermodynamic stability of these black holes.

However,  this
Regge-Wheeler
prescription is limited to the second-order gravity and thus, it  is no longer  applicable to a higher-order gravity.
For a fourth-order gravity, the linearized Ricci tensor  $\delta R_{\mu\nu}$  may represent  a   massive spin-2 mode, while
the linearized Ricci scalar  $\delta R$ denotes  a   massive spin-0 mode~\cite{Whitt:1985ki,Mauro:2015waa,Stelle:2017bdu}.
A merit of  introducing  these variables is to reduce a fourth-order  gravity to a second-order  gravity without  introducing  ghosts.
It was  shown that the linearized Ricci tensor theory exhibits unstable modes featuring the Gregory-Laflamme (GL) instability in Ricci quadratic gravity (Einstein-Weyl gravity).
Actually, this was
performed by comparing the linearized Ricci tensor equation  with
the linearized metric tensor equation around the five-dimensional black string where the GL instability appeared~\cite{Gregory:1993vy}.

Interestingly, a close connection between
classical and  thermodynamic instability for the black strings/branes in the metric tensor formalism was proposed by considering  Reissner-Nordstr\"{o}m-AdS black hole in the Kaluza-Klein reduction of the  ${\cal N}=8$ gauged supergravity~\cite{Gubser:2000ec,Gubser:2000mm}.  This Gubser-
Mitra proposal was known to be the correlated stability conjecture (CSC)~\cite{Harmark:2007md}.
This conjecture  states that {\it  gravitational systems with translational symmetry and infinite extent exhibit GL instability if and only if they have a local thermodynamic instability}.   It is easily checked that the CSC does not hold for Schwarzschild and SAdS black holes obtained from GR because they have not  translational symmetry
and infinite extent. However,
it was found that the CSC holds for the $n(\ge 4)$-dimensional AdS black hole in Einstein-Weyl  gravity  by establishing a close relation between GL instability and thermodynamic instability~\cite{Myung:2013uka}.
Also, the CSC holds for SAdS black hole in Einstein-Ricci cubic gravity~\cite{Myung:2018ete}.
 In these cases, one notes that  the massiveness of a massive spin-2 mode takes over a role of the higher-dimensional black sting.

Furthermore, a newly non-Schwarzschild black hole solution has been found  from  the linearized Ricci quadratic gravity by solving   static Lichnerowicz equations~\cite{Lu:2017kzi}. This static solution plays two roles of  a perturbation away from the  Schwarzschild black hole along the newly non-Schwarzschild black hole and   a threshold unstable mode lying at the edge of a domain of GL instability for small Schwarzschild black holes.  This confirms the threshold mass for GL instability  in Ricci quadratic gravity.

It is interesting to introduce  another higher-derivative theory of  six-order gravity.
One six-order gravity is Einsteinian cubic gravity including  Riemann cubic polynomial ${\cal P}_3$, which  was shown to be  neither topological nor trivial in four dimensions~\cite{Bueno:2016xff}.
Even though black hole solutions to this gravity have interesting properties~\cite{Hennigar:2017ego,Ahmed:2017jod,Bueno:2016lrh},  these belong to either  numerical or approximate solutions. This means that the Einstein equation  cannot be solved analytically and  an obstacle to study these black holes is a lack of  an analytic solution.
The other sixth-order gravity is given by the Ricci cubic gravity~\cite{Li:2017ncu}. It includes  Ricci cubic polynomial ${\cal R}_3$. We note that   ${\cal R}_3$ is  much more manageable than  ${\cal P}_3$. A benefit of Ricci cubic  gravity is that the $n$-dimensional AdS black hole solution  to Einstein gravity is also a solution to this gravity. In addition, one is able  to  construct a covariant linearized theory on the AdS$_n$ black hole in Ricci cubic gravity, whereas the covariant linearized theory of Einsteinian cubic gravity is allowed only on a vacuum  spacetime of AdS$_n$.  These are reasons why we  choose Ricci cubic gravity for the study of AdS black holes.

In this work, we will investigate  classical (GL) and thermodynamic instability for $n$-dimensional  AdS black
holes in Ricci cubic gravity.  After obtaining a covariant linearized theory on the $n$-dimensional AdS black hole background,  we show that  a small black hole with $r_+<r_*^{(n)}=\sqrt{\frac{n-3}{n-1}}\ell$ is unstable against the Ricci tensor perturbation, whereas a large black hole with $r_+>r_*^{(n)}$ is stable.
Then,  computing the Wald entropy, we derive the heat capcity in $n$-dimensional Ricci cubic  gravity.  Hence, we will establish a close relation between  the GL and thermodynamic instability for AdS black holes. Finally, we wish to find a newly non-AdS black hole
by solving    the  static Lichnerowicz equation. This case confirms  the existence of the threshold mass $M^t_n$ defined an edge of GL instability of AdS black holes.

\section{$n$-dimensional Ricci cubic gravity} \label{sec1}

We start with the Ricci cubic  gravity  in $n(\ge4)$-dimensional  spacetimes~\cite{Li:2017ncu}
\begin{eqnarray}
S_{\rm RC}\equiv\frac{1}{16 \pi}\int d^n x\sqrt{-g}{\cal L}_{\rm RC}=\frac{1}{16 \pi}\int d^n x\sqrt{-g}[\kappa_n(R-2\Lambda_0)+{\cal R}_3],\label{Action1}
\end{eqnarray}
where  Ricci cubic polynomials ${\cal R}_3$ are given by
\begin{eqnarray}\label{Action2}
{\cal R}_3=e_1R^3+e_2 R R_{\mu\nu}R^{\mu\nu}+e_3 R^\mu_\nu R^\nu_\rho R^\rho_\mu.
\end{eqnarray}
Here $\kappa_n=1/G_n$ is the inverse of $n$-dimensional Newtonian constant, $\Lambda_0$ is the bare cosmological constant, and  $\{e_1,e_2,e_3\}$  denote three cubic  parameters to specify Ricci polynomials ${\cal R}_3$. Their mass dimensions are given by $[\kappa_n]=n-2$, $[\Lambda_0]=2$, and $[ e_1,e_2,e_3]=n-6$.

From the action (\ref{Action1}), we derive  the Einstein equation as
\begin{eqnarray}
&&P_{\mu\alpha\beta\gamma}R_\nu~^{\alpha\beta\gamma}
-2\nabla^\alpha\nabla^\beta P_{\mu\alpha\beta\nu}-\frac{1}{2}g_{\mu\nu}{\cal L}_{\rm RC} \label{ein-eq}=0,  \label{equa1}
\end{eqnarray}
where the $P$-tensor  is given by
\begin{eqnarray}
P_{\mu\nu\rho\sigma}&\equiv&\frac{\partial {\cal L}_{\rm RC}}{\partial R^{\mu\nu\rho\sigma}}= \frac{\kappa_n}{2} (g_{\mu\rho}g_{\nu\sigma}-g_{\mu\sigma}g_{\nu\rho})+\frac{3e_1}{2}R^2(g_{\mu\rho}g_{\nu\sigma}-g_{\mu\sigma}g_{\nu\rho})\nonumber \\
&+&\frac{e_2}{2}R_{\alpha\beta}R^{\alpha\beta}(g_{\mu\rho}g_{\nu\sigma}-g_{\mu\sigma}g_{\nu\rho})+\frac{e_2}{2}R(g_{\mu\rho}R_{\nu\sigma}-g_{\mu\sigma}R_{\nu\rho}
-g_{\nu\rho}R_{\mu\sigma}+g_{\nu\sigma}R_{\mu\rho})\nonumber \\
&+&\frac{3e_3}{4}(g_{\mu\rho}R_{\nu\gamma}R^\gamma_\sigma-g_{\mu\sigma}R_{\nu\gamma}R^\gamma_\rho-g_{\nu\rho}R_{\mu\gamma}R^\gamma_\sigma+g_{\nu\sigma}R_{\mu\gamma}R^\gamma_\rho).  \label{equa2}
\end{eqnarray}
For a static black hole with spherical symmetry, the metric has the form
\begin{equation} \label{ansatz}
ds^2=-h(r)dt^2+\frac{dr^2}{f(r)}+r^2d\Omega^2_{n-2}.
\end{equation}
Notice that an AdS black
hole solution obtained from the $n$-dimensional Einstein gravity
is also a solution to Eq.(\ref{equa1}), whose form  is given by
\begin{equation} \label{num}
h(r)=f(r)=\bar{f}(r)=1-\Big(\frac{r_0}{r}\Big)^{n-3}-\frac{a}{n-1}r^2,~a=\frac{2\Lambda}{n-2}=-\frac{n-1}{\ell^2}
\end{equation}
with $\ell$  the
curvature radius of AdS$_n$ spacetimes. Hereafter we denote all background quantities (AdS black hole solutions) with the overbar as
\begin{equation}\label{sch}
ds^2_{\rm AdS}\equiv\bar{g}_{\mu\nu}dx^\mu dx^\nu=-\bar{f}(r)dt^2+\frac{dr^2}{\bar{f}(r)}+r^2d\Omega^2_{n-2}.
\end{equation}
A black hole mass
parameter $r_0$ is determined to be
\begin{equation} \label{ro-eq}
r_0=\Big[r_+^{n-3}+\frac{r_+^{n-1}}{\ell^2}\Big]^{\frac{1}{n-3}}
\end{equation}
which  is not exactly  the horizon radius
$r_+$.  The background Ricci tensor and Ricci scalar  are determined  by
\begin{equation} \label{beeq}
\bar{R}_{\mu\nu}= a\bar{g}_{\mu\nu} ,~\bar{R}=na.
\end{equation}
Plugging (\ref{beeq}) into (\ref{equa2}),  the background
 $P$-tensor takes a maximally symmetric form as
\begin{eqnarray}
\bar{P}_{\mu\nu\rho\sigma}&=&\frac{1}{2}\Big[\kappa_n+3(n^2e_1+ne_2+e_3)a^2\Big](\bar{g}_{\mu\rho}\bar{g}_{\nu\sigma}-\bar{g}_{\mu\sigma}\bar{g}_{\nu\rho})\nonumber \\
&=&\frac{1}{2}\Big[\kappa_n+(\alpha+n\beta)a^2\Big](\bar{g}_{\mu\rho}\bar{g}_{\nu\sigma}-\bar{g}_{\mu\sigma}\bar{g}_{\nu\rho})\label{p-tensor}
\end{eqnarray}
with introducing two new parameters
\begin{equation}
\alpha=ne_2+3e_3,~\beta=3ne_1+2e_2.
\end{equation}
The two parameters $\alpha$ and $\beta$ are enough to denote the Ricci polynomial ${\cal R}_3$ on the AdS black hole background, instead of three parameters $\{e_1,e_2,e_3\}$.
The expression (\ref{p-tensor}) will  be used to derive the Wald entropy in section 5.
Here, the effective cosmological constant $\Lambda$ is related to the bare cosmological constant $\Lambda_0$ as
\begin{equation}
\Lambda+\frac{4(n-6)}{3\kappa_n(n-2)^3}(\alpha+n\beta)\Lambda^3= \Lambda_0.
\end{equation}
In the case of $\alpha=\beta=0(e_2=-\frac{3ne_1}{2},~e_3=\frac{n^2e_1}{2})$, one finds $ \Lambda= \Lambda_0$ in quasi-topological gravity.
It is easy to show that the AdS black hole  (\ref{sch})
to  Einstein equation of $G_{\mu\nu}=\Lambda\bar{g}_{\mu\nu}$  is also the solution
to the Ricci cubic  gravity when one substitutes (\ref{p-tensor})  together with (\ref{beeq}) into (\ref{equa1}).
However, the background Riemann tensor for the AdS black hole is not given by the  AdS$_n$-curvature tensor
\begin{equation}
\bar{R}_{\mu\nu\rho\sigma}\not=\bar{R}_{\mu\nu\rho\sigma}^{\rm AdS_n}=\frac{a}{n-1}(\bar{g}_{\mu\rho}\bar{g}_{\nu\sigma}-\bar{g}_{\mu\sigma}\bar{g}_{\nu\rho}),
\end{equation}
which implies  that the AdS black hole  (\ref{sch}) is not a maximally symmetric vacuum.

\section{Linearized theory: Lichnerowicz equation}

In order to obtain a linearized theory defined around AdS black holes, let us  introduce the metric
perturbation around  AdS black holes as
\begin{eqnarray} \label{m-p}
g_{\mu\nu}=\bar{g}_{\mu\nu}+h_{\mu\nu}.
\end{eqnarray}
The linear stability of the black hole solution (\ref{sch}) to  Eq.(\ref{equa1}) is usually sufficient for guaranteeing the stability at  any perturbative  level.
According to Refs.~\cite{Whitt:1985ki,Stelle:2017bdu}, it is desirable  to investigate  the stability of AdS black hole by means of second-order equation for the linearized Ricci tensor, instead of  fourth-order equation for the metric perturbation.
Therefore, we  define the linearized  Ricci tensor and scalar  as
\begin{equation}
\delta \tilde{R}_{\mu\nu}=\delta R_{\mu\nu}-a h_{\mu\nu},~\delta R=\delta(g^{\mu\nu}R_{\mu\nu})=\bar{g}^{\mu\nu}\delta \tilde{R}_{\mu\nu},
\end{equation}
where the conventionally linearized Ricci tensor and scalar are given by
\begin{eqnarray}
\delta R_{\mu\nu}(h)&=&\frac{1}{2}\Big[\bar{\nabla}^{\rho}\bar{\nabla}_{\mu}h_{\nu\rho}+
\bar{\nabla}^{\rho}\bar{\nabla}_{\nu}h_{\mu\rho}-\bar{\square}h_{\mu\nu}-\bar{\nabla}_{\mu}
\bar{\nabla}_{\nu}h\Big], \label{ricc-t} \\
\delta R(h)&=&\bar{\nabla}^\mu
\bar{\nabla}^\nu h_{\mu\nu}-\bar{\square} h-a h
\label{Ricc-s}
\end{eqnarray}
with $h=h^\rho_\rho$.
Roughly speaking, the linearized Ricci tensor represents
a healthy massive spin-2 mode, while the linearized Ricci scalar represents
a healthy massive spin-0 mode.
Conveniently, we may introduce  the linearized Einstein tensor as
\begin{equation}
\delta G_{\mu\nu}=\delta \tilde{R}_{\mu\nu}-\frac{1}{2}\bar{g}_{\mu\nu}\delta R=\delta R_{\mu\nu}-\frac{1}{2}\bar{g}_{\mu\nu}\delta R-ah_{\mu\nu}, \label{LET}
\end{equation}
which is  the linearization of the Einstein tensor
\begin{equation}
G_{\mu\nu}=R_{\mu\nu}-\frac{R}{2}g_{\mu\nu}+\frac{(n-2)a}{2}g_{\mu\nu}.
\end{equation}
Linearizing Eq.(\ref{equa1}) leads to the linearized Einstein equation
\begin{eqnarray}
&&(\kappa_n-a\alpha\bar{\Delta}_{\rm L}) \delta G_{\mu\nu}+a^2(3\alpha+n\beta)\delta \tilde{R}_{\mu\nu}
-\frac{1}{2}a^2 [\alpha+(n-4)\beta]\bar{g}_{\mu\nu} \delta R \nonumber \\
&&-a (\alpha+2\beta)(\bar{\nabla}_\mu \bar{\nabla}_\nu-\bar{g}_{\mu\nu}\bar{\square} )\delta R =0, \label{l-eineq}
\end{eqnarray}
where the Lichnerowicz operator is  defined  by acting  on linearized Ricci scalar and tensor, respectively,
\begin{eqnarray}
&&\bar{\Delta}_{\rm L}\delta R=-\bar{\square}\delta R, \label{lich1} \\
&& \bar{\Delta}_{\rm L} \delta \tilde{R}_{\mu\nu}=-\bar{\square}\delta \tilde{R}_{\mu\nu}-2 \bar{R}^\rho~_{\mu \sigma \nu}\delta \tilde{R}_{\rho}^\sigma+\bar{R}_\mu^\rho \delta \tilde{R}_{\rho\nu}
+\bar{R}_\nu^\rho \delta \tilde{R}_{\rho\mu}. \label{lich2}
\end{eqnarray}
The linearized equation (\ref{l-eineq}) is  a second-order equation for $\delta R_{\mu\nu}$, but  it becomes a fourth-order equation when expressing  in terms of  $h_{\mu\nu}$.
Putting $a=0(\Lambda=0)$ yields Ricci-flat solution (Schwarzschild solution) on which Ricci polynomial  gives no contributions to the linearized equation, leading to
$\delta R_{\mu\nu}=0$.
This is one reason why we included the  cosmological constant in the beginning action (\ref{Action1}), compared to the fourth-order gravity.
Taking the trace of Eq.(\ref{l-eineq})  leads to the linearized Ricci scalar equation
\begin{equation}
a[n\alpha+4(n-1)\beta]\bar{\square}\delta R-[(n-2)\kappa_n+(n-6)(\alpha+n\beta)a^2] \delta R=0.\label{LE-0}
\end{equation}
One notes that Eq.(\ref{l-eineq}) is a coupled second-order equation for $\delta \tilde{R}_{\mu\nu}$ and $\delta R$, which seems difficult to be solved.
One way to avoid this difficulty is to split Eq.(\ref{l-eineq}) into the traceless and trace parts by choosing $\alpha$ appropriately.
For this purpose, we introduce a traceless Ricci tensor as
\begin{equation} \label{hatR}
 \delta\hat{ R}_{\mu\nu}= \delta \tilde{R}_{\mu\nu}-\frac{1}{n}\bar{g}_{\mu\nu} \delta R,~~ \delta \hat{R}=0.
\end{equation}
Then, Eqs.(\ref{l-eineq}) and (\ref{LE-0}) lead to
\begin{eqnarray}
&& \alpha (\bar{\Delta}_{\rm L} -2a+\mu^2_2)\delta \hat{ R}_{\mu\nu}+ (\alpha+2\beta) \Big(\bar{\nabla}_\mu \bar{\nabla}_\nu-\frac{1}{n}\bar{g}_{\mu\nu}\bar{\square}\Big)\delta R=0,\label{LE-1}\\
&&(\bar{\square}-\mu_0^2) \delta R=0, \label{LE-2}
\end{eqnarray}
where the  mass squared $\mu^2_2$ for massive spin-2 mode and  mass squared $\mu^2_0$  for massive   spin-0 mode  are given by
\begin{eqnarray}
 \mu_2^2&=&2a-\frac{(3\alpha+n\beta)a^2+\kappa_n}{a \alpha},~~\mu^2_0=\frac{(n-2)\kappa_n+(n-6)(\alpha+n\beta)a^2}{a[n\alpha+4(n-1)\beta]}. \label{masses}
\end{eqnarray}
We note that decoupling of all massive modes requires either $a=0$ or $ \alpha=\beta=0$.  The former case corresponds to the Ricci-flat spacetimes on which Ricci polynomial gives no contribution to the linearized Ricci tensor equation.
On the other hand, the latter case yields a quasi-topological gravity,
\begin{equation} \label{QTG}
{\cal L}_{\rm RC}^{\alpha=\beta=0}={\cal L}_{\rm QT}=\kappa_n(R-2\Lambda_0)+e_1\Big(R^3-\frac{3n}{2}RR_{\mu\nu}R^{\mu\nu}+\frac{n^2}{2}R^\mu_\nu R^\nu_\rho R^\rho_\mu\Big)
\end{equation}
whose linearized equation is exactly given by~\cite{Li:2017ncu}
\begin{equation} \label{lQTG}
\delta G_{\mu\nu}=0.
\end{equation}
Hence, it leads to a ghost-free gravity when perturbing around the Minkowski spacetimes.
Also, the AdS black hole found from this theory is stable against metric perturbations.

When choosing a condition of  $\alpha=-2\beta$, a nontrivial decoupling between traceless and trace parts occurs naturally.
This case implies that  one parameter $e_3$  is redundant  and it is represented by
\begin{equation}
e_3=-2ne_1-\frac{(n+4)e_2}{3}.
\end{equation}
In this case, Eqs.(\ref{LE-1}) and (\ref{LE-2}) lead to the massive spin-2 and massive spin-0 equations, separately,
\begin{eqnarray}
&&(\bar{\Delta}_{\rm L} -2a+M^2_n)\delta \hat{ R}_{\mu\nu}=0, \label{LE-3} \\
&&(\bar{\square}-M_0^2) \delta R=0. \label{LE-4}
\end{eqnarray}
Here the mass squared $M^2_n$ and $M^2_0$ are given by
\begin{equation}
M^2_n=\frac{\kappa_n\ell^2}{(n-1)\alpha}-\frac{(n-1)(n-2)}{2\ell^2},~M^2_0=\frac{\kappa_n\ell^2}{(n-1)\alpha}-\frac{3(n-6)}{2\ell^2}.
\end{equation}
However, the number of  degrees of freedom (DOF) for $\delta \hat{ R}_{\mu\nu}$  is not given by  $(n+1)(n-2)/2$ for a  massive spin-2 mode in $n$ dimensions because the contracted Bianchi identity ($\bar{\nabla}^\mu\delta G_{\mu\nu}=0$) does not imply
the transverse condition as
\begin{equation}
\bar{\nabla}^\mu\delta \hat{R}_{\mu\nu}=\Big[\frac{n-2}{2n}\Big]\bar{g}_{\mu\nu}\bar{\nabla}_\nu \delta R \nrightarrow \bar{\nabla}^\mu\delta \hat{R}_{\mu\nu}=0
\end{equation}
because   $\delta R\not=0$ in the $\alpha=-2\beta$ Ricci cubic gravity.  This is not the case.

A desirable choice may be  achieved  by requiring  a non-propagation of the linearized Ricci scalar ($\delta R=0$).
Imposing a condition of $\alpha=-4(n-1)\beta/n$ on the linearized  Ricci scalar equation (\ref{LE-0}), the non-propagating Ricci scalar is achieved   as
\begin{equation}
\delta R=0. \label{MLE-0}
\end{equation}
In other words, when  choosing
\begin{equation}
e_3=-4(n-1)\Big(e_1+\frac{2e_2}{3n}\Big),
\end{equation}
one could achieve the non-propagation of linearized Ricci scalar.
In this case, the theory is given by
\begin{equation} \label{spec}
{\cal L}_{\rm RC}^{\alpha=-4(n-1)\beta/n}=\kappa_n(R-2\Lambda_0)+e_1R^3+e_2RR_{\mu\nu}R^{\mu\nu}-4(n-1)\Big(e_1+\frac{2e_2}{3n}\Big)R^\mu_\nu R^\nu_\rho R^\rho_\mu.
\end{equation}
Also, one finds from (\ref{masses}) that the  mass squared $\mu^2_0$ of a massive spin-0 mode blows up at $\alpha=-4(n-1)\beta/n$, which means that the massive spin-0 mode is decoupled from the theory. The absence of massive spin-0 mode is also required by an $a$-theorem~\cite{Li:2017txk}. In addition, the ghost-free condition may be obtained when imposing  the absence of massive spin-2 mode. The absence of both massive modes are required by the causality condition such  that the resulting theory becomes a linearized quasi-topological theory (\ref{lQTG}).
\begin{figure*}[t!]
   \centering
   \includegraphics{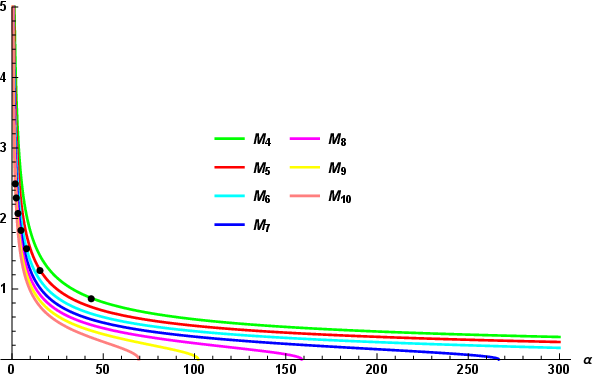}
\caption{Plots of mass $M_n$ for a massive spin-2 mode as a function of $\alpha$  with
$l=10$ and $\kappa_n=1$ in the $\alpha=-4(n-1)\beta/n$  Ricci cubic  gravity.
 For positive $M^2_n$, there is no tachyonic instability but the Gregory-Laflamme instability will appear for $0<M_n<M^t_n$ with $M^t_n$ threshold mass. `$\bullet$' denotes ($\alpha^t_n,M^t_n$)  in (\ref{cwn}).  $M_n>M^t_n$ correspond to the stable cases. }
\end{figure*}
Going back to the classical stability analysis of AdS black holes, we choose  $\delta R=0$.
The ghost-problem is not present here because  one adopts  $\delta \hat{R}_{\mu\nu}$ to represent a massive spin-2 mode, instead of $h_{\mu\nu}$.
Considering $\delta G_{\mu\nu}$ in (\ref{LET}) together with $\delta R=0$, Eq.(\ref{l-eineq}) leads to
a massive spin-2 equation for $\delta \hat{R}_{\mu\nu}$ (so-called Lichnerowicz equation) as
\begin{equation}
(\bar{\Delta}_{\rm L} -2a+M^2_n)\delta \hat{ R}_{\mu\nu}=0 \label{LE-5}
\end{equation}
with the mass
\begin{equation}
M_n=\sqrt{\frac{\kappa_n\ell^2}{(n-1)\alpha}-\frac{(n-2)^2}{4\ell^2}}. \label{MLE-1}
\end{equation}

Fig. 1 depicts the mass $M_n$ for a massive spin-2 mode as a function of $\alpha$ and $n$. $M_n=0$ appears at $\alpha=\alpha_n=4\kappa_n\ell^4/(n-1)(n-2)^2(\alpha_4=3333.3,~\alpha_5=1111.1,~\alpha_6=500,~\alpha_7=266.7,~\alpha_8=158.3,\alpha_9=102,~\alpha_{10}=69.4)$.
$\alpha_n$ decreases as $n$ increases. We observe that  there is no distinct features depending on dimension $n$.

 In the AdS black hole background, a simple criterion of the stability for massive spin-2 mode
requires  the positive mass squared of $M^2_n> 0$, which implies that
\begin{equation} \label{alpha-c}
0<\alpha< \alpha_n.
\end{equation}
This  corresponds to  tachyon-free condition for $\delta \hat{R}_{\mu\nu}$ propagating on the AdS black hole background~\cite{Liu:2011kf}.  However, $M^2_n> 0~(0<\alpha<\alpha_n)$ is not a necessary and sufficient condition for getting a stable AdS black hole. One needs a further analysis to complete the stability analysis of the AdS black hole.  In the next section, the Gregory-Laflamme (GL) instability will appear for $0<M_n<M^t_n$ with $M^t_n$ the threshold mass for GL instability.  If $M^2_n<0(\alpha>\alpha_n)$, one does not need to perform  a further analysis because it
implies the tachyonic instability. Interestingly, the case of $M_n^2=0(\alpha=\alpha_n)$ indicates  a critical gravity when one employs the transverse-traceless gauge for metric perturbation $h_{\mu\nu}$, leading to another instability of log-gravity.

Finally, considering  Eq.(\ref{MLE-0}), the contracted Bianchi identity yields  a desired  transverse condition as
\begin{equation}
\bar{\nabla}^\mu\delta \hat{R}_{\mu\nu}=0. \label{MLE-3}
\end{equation}
Hence, the DOF of $\delta \hat{R}_{\mu\nu}$ becomes $(n+1)(n-2)/2$ from counting of  $(n+1)(n+2)/2-n-1$ in the  $\alpha=-4(n-1)\beta/n$ Ricci cubic gravity. From now on, we consider the  $\alpha=-4(n-1)\beta/n$ Ricci cubic gravity in Eq.(\ref{spec}) only.

\section{GL instability for AdS black holes}

When performing  the stability analysis of the AdS black hole in $\alpha=\beta=0$ Ricci cubic
gravity (quasi-topological gravity)~\cite{Li:2017ncu}, one has to use  the linearized equation (\ref{lQTG}) designed for Einstein gravity.
In this case, it turned  out to be stable by making use of  the Regge-Wheeler
prescription~\cite{Cardoso:2001bb,Ishibashi:2003ap,Moon:2011sz}. Here, considering  the $s(l=0)$-mode  is unnecessary to show the
stability of   AdS black holes because a massless spin-2 mode starts from $l=2$.

Now, let us consider the linearized $\alpha=-4(n-1)\beta/n$ Ricci cubic gravity.  Its  Lichnerowicz equation (\ref{LE-5}) is second-order  with respect to $\delta \hat{R}_{\mu\nu}$.
Importantly, the traceless Ricci tensor $\delta\hat{ R}_{\mu\nu}$ representing a massive spin-2 mode  satisfies the transverse condition (\ref{MLE-3}).
This means that the transverse-traceless (TT)  gauge condition  ($\bar{\nabla}^\mu\delta \hat{R}_{\mu\nu}=0$ and  $\delta \hat{R}=0$) is allowed for $\delta\hat{ R}_{\mu\nu}$.
Hence, its DOF is determined to be $(n+1)(n-2)/2$ as for $\delta G_{\mu\nu}$ in the $n$-dimensional  Einstein-Weyl gravity.
At this stage, the even-parity metric perturbation is necessary for a single $s$-mode analysis of $\delta \hat{R}_{\mu\nu}$
whose form is given by~\cite{Gregory:1994bj}
\begin{eqnarray}
\delta \hat{R}_{\mu\nu}(t,r,\cdots)=e^{\Omega t} \left(
\begin{array}{cccccc}
H_{ij}(r)& H_{it}(r)& H_{ir}(r)&0&0&\cdots \cr
H_{tj}(r)& H_{tt}(r) & H_{tr}(r) & 0 & 0 & \cdots \cr
H_{rj}(r)&  H_{tr}(r) &H_{rr}(r) & 0 & 0& \cdots \cr
0&0 & 0 & K(r) & 0&\cdots \cr
0&0 & 0 & 0 & \sin^2\theta K(r)&\cdots \cr
\cdots&\cdots&\cdots&\cdots&\cdots&\cdots
\end{array}
\right) \label{evenp}
\end{eqnarray}
with $i=5, \cdots,n$. It is suggested  that $\delta \hat{R}_{\mu\nu}$ grows exponentially in time as $e^{\Omega t}$ with  $\Omega>0$, while it spatially vanishes at the AdS infinity and
it is regular at the horizon.

Starting  with 3 DOF ($H_{tt},~H_{tr},~H_{rr}$),  they are  related to each other
when imposing   the TT gauge condition.
In order to investigate the classical instability for
$n$-dimensional AdS black holes, we  define the new perturbed variables as
\begin{eqnarray}
H\equiv H_{tr},~~~H_\pm\equiv\frac{H_{tt}}{\bar{f}(r)}\pm
\bar{f}(r)H_{rr}.
\end{eqnarray}
Then, one finds  two coupled
first-order equations as
\begin{eqnarray}
H'&=&\Big[\frac{3-n-(n-1)r^2/\ell^2}{r\bar{f}}-\frac{1}{r}\Big]H+\frac{\Omega}{2\bar{f}}(H_++H_-),\label{Hsd}\\
H_{-}^{\prime}&=&\frac{M_n^2}{\Omega}H+\frac{n-2}{2r}H_++\Big[\frac{n-3+(n-1)
r^2/\ell^2}{2r\bar{f}}-\frac{2n-3}{2r}\Big]H_-.\label{Hsmd}
\end{eqnarray}
In addition,  a constraint equation is given by
\begin{eqnarray}
&&r^2\Omega\Big[4r\Omega^2-r(\bar{f}')^2+(n-2)\bar{f} \bar{f}'+2r\bar{f}M_n^2+2r\bar{f} \bar{f}''\Big]H_-\nonumber\\
&&-\Omega r^2\bar{f}\Big[2
M_n^2r+(n-2)\bar{f}'\Big]H_+ -2r^2\bar{f}\Big[2(n-2)\Omega^2-2M_n^2\bar{f}+rM_n^2\bar{f}'\Big]H=0.\label{cons}
\end{eqnarray}
We note that two first-order equations and one constraint equation can be recovered from
 second-order linearized equation (\ref{LE-5}) together with  the  TT gauge
condition.

 By eliminating $H_+$ in Eqs.(\ref{Hsd}) and (\ref{Hsmd}) together with the constraint (\ref{cons}),
one finds two coupled equations for  $H$ and  $H_-$. Firstly, we wish to find  an asymptotic solution.
At the AdS infinity of $r\to\infty$, the  asymptotic solution is  given by
\begin{eqnarray}
H^{(\infty)}&=&c^{(\infty)}_{1}r^{-(n+1)/2+\sqrt{M_n^2\ell^2+(n-1)^2/4}} +c^{(\infty)}_{2}r^{-(n+1)/2-\sqrt{M_n^2\ell^2+(n-1)^2/4}},\nonumber\\
&&\nonumber\\
H_-^{(\infty)}&=&\tilde{c}^{(\infty)}_{1}r^{-(n-1)/2+\sqrt{M_n^2\ell^2+(n-1)^2/4}}
+\tilde{c}^{(\infty)}_{2}r^{-(n-1)/2-\sqrt{M_n^2\ell^2+(n-1)^2/4}},
\end{eqnarray}
where coefficients $\tilde{c}^{(\infty)}_{1/2}$ take the forms
\begin{eqnarray}
\tilde{c}^{(\infty)}_{1/2}&=&\frac{M_n^2}{(1-n)/2\pm\sqrt{M_n^2\ell^2+(n-1)^2/4}}{c}^{(\infty)}_{1/2}.
\end{eqnarray}
 At the horizon
$r=r_+$, their  solution is given by
\begin{eqnarray}
H^{(r_+)}&=&c^{(r_+)}_{1}(r^{n-3}-r_+^{n-3})^{-1+\frac{\Omega}{\bar{f}'(r_+)}}
+c^{(r_+)}_{2}(r^{n-3}-r_+^{n-3})^{-1-\frac{\Omega}{\bar{f}'(r_+)}},\nonumber\\
&&\nonumber\\
H_-^{(r_+)}&=&\tilde{c}^{(r_+)}_{1}(r^{n-3}-r_+^{n-3})^{\frac{\Omega}{\bar{f}'(r_+)}}
+\tilde{c}^{(r_+)}_{2}(r^{n-3}-r_+^{n-3})^{-\frac{\Omega}{\bar{f}'(r_+)}}.
\end{eqnarray}
Here, coefficients  $\tilde{c}^{(r_+)}_{1/2}$ are given by
\begin{eqnarray}
\tilde{c}^{(r_+)}_{1/2}&=&\pm\frac{(n-3)r_+^{n-3}\Omega\Big(2\Omega \mp \bar{f}'(r_+)\Big)}{2\bar{f}'(r_+)\Big(M_n^2r_+\pm(n-2)\Omega\Big)}~{c}^{(r_+)}_{1/2}.
\end{eqnarray}
We note that two boundary conditions for obtaining  regular solutions
correspond to ${c}^{(\infty)}_{1}=0$ at
infinity and ${c}^{(r_+)}_{2}=0$ at the horizon.
For given dimension $n \in [4,\cdots,10]$ and fixed
$M_n$,  we solve two coupled
equations numerically.  They yield allowed values of $\Omega$ as functions of $M_n$.
Fig. 2 shows  that  the curves of $\Omega$ intersect the  positive $M_n$-axis at one place: $M_n=M^t_n$ where  $M_n^t$ is
a threshold mass for GL instability.
\begin{figure*}[t!]
   \centering
   \includegraphics{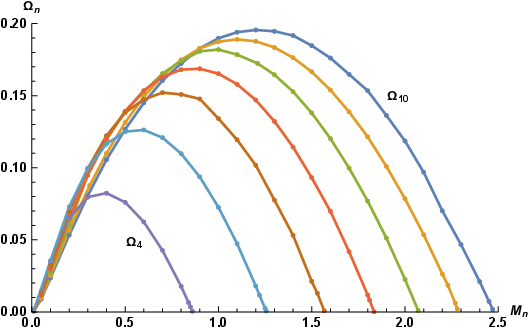}
\caption{$\Omega_n$ graphs as function of $M_n$ for  small
AdS black holes with $r_+(=1)<r_*,~\ell=10$ and $n=4,5,\cdots,10$ from left to
right curve. Here one may read off the threshold mass  $M_n^t$ from the points that curves of $\Omega_n$ cross the positive $M_n$-axis.  }
\end{figure*}
By observing  Fig. 2, we read off the
threshold mass $M_n^t$ depending on the dimension $n$ as
\begin{equation} \label{cwn}
\left[
\begin{array}{c|ccccccc}
 n & 4 & 5 & 6 & 7 & 8 & 9 & 10 \cr \hline
      M_n^t& 0.86 & 1.26 & 1.57 & 1.83 & 2.07 & 2.29& 2.49 \cr \hline
       \alpha_n^t& 43.5& 15.3 & 7.89 &4.89 & 3.27& 2.33&1.75 \cr
\end{array}
\right].
\end{equation}
Here, $\alpha_n^t$ is determined by $M_n^t$ through Eq.(\ref{MLE-1}).
It is important to note that for all large AdS  black holes of $r_+>r_*^{(n)}=\sqrt{\frac{n-3}{n-1}}\ell$,   $\Omega$ approaches the maximum value being less than $10^{-4}$.
This implies that there is no unstable modes for large AdS black holes  in $\alpha=-4(n-1)\beta/n$-Ricci cubic gravity.
The other crossing  points at the origin ($M_n=0,\alpha=\alpha_n$) mean that AdS black holes  are unstable in the critical theory of Ricci cubic gravity~\cite{Liu:2011kf}.
For a given curve,  $0<M_n<M_n^t(M_n>M_n^t)$ implies that the
AdS black holes are unstable (stable) against the Ricci tensor perturbation.
Observing Fig. 1, one easily finds that  such unstable (stable) regions are always allowed.

Now we are in a position to find the GL instability condition explicitly.
For $r_+=1$ and $\ell=10$,  the mass parameter  takes the form  from Eq.(\ref{ro-eq})
\begin{equation}
r_0=\Big(\frac{101}{100}\Big)^{\frac{1}{n-3}},
\end{equation}
which implies that  $r_0=\{1.01,1.005,1.003,1.002,1.002,1.002,1.001\}$.
Here, the corresponding wave number $\tilde{k}_n^t=r_0M_n^t$  is determined to be~\cite{Kol:2004pn,Kol:2006vu,Held:2022abx}
\begin{equation} \label{kwn}
\left[
\begin{array}{c|ccccccc}
 n & 4 & 5 & 6 & 7 & 8 & 9 & 10 \cr \hline
      \tilde{k}^t_n& 0.87 & 1.27 & 1.57 & 1.83 & 2.07 & 2.29& 2.49 \cr
\end{array}
\right].
\end{equation}
Hence, we propose the bound for unstable modes  as
\begin{equation}
0<M_n<\frac{\tilde{k}_n^t}{r_0}=M^t_n,\label{mass-b}
\end{equation}
which  indicates  that  the GL
instability of small AdS black holes in the $\alpha=-4(n-1)\beta/n$ Ricci cubic gravity is due to
the massiveness of $M_n \not=0$. The ghost  may appear only when expressing the linearized equation (\ref{LE-5}) in terms of the  metric perturbation $h_{\mu\nu}$.
It is worth noting that  the ghost (unhealthy  massive spin-2 mode) does not appear here because we adopt  the linearized Ricci tensor $\delta \hat{R}_{\mu\nu}$ to represent a massive spin-2 mode.

Finally, it is well known that the CSC proposed by Gubser-Mitra~\cite{Gubser:2000ec}  does not hold for  SAdS black hole obtained
 from  Einstein gravity, but it holds for  SAdS black hole found from Ricci quadratic gravity (Einstein-Weyl  gravity)~\cite{Myung:2013uka} and from Einstein-Ricci cubic gravity~\cite{Myung:2018ete}.
 In this direction, it suggests that  the GL instability
condition (massiveness) picked up  small AdS black holes with
$r_+<r^{(n)}_*$ which may be thermodynamically unstable in Ricci cubic gravity.
Hence, it is desirable to explore a deep  connection between the
GL and thermodynamic instability of   AdS black
holes in Ricci cubic  gravity.

\section{Thermodynamic instability}
First of all, all
thermodynamic quantities (mass, heat capacity, Bekenstein-Hawking entropy, Hawking temperature, Helmholtz free energy)
of AdS black holes in $n$-dimensional Einstein gravity with a cosmological constant $\Lambda$
were known to be~\cite{Myung:2013uka}
\begin{eqnarray} \label{hbtz1}
m_n(r_+)&=&\frac{\Omega_{n-2}(n-2)}{16\pi
G_n}r_+^{n-3}\Big[1+\frac{r_+^2}{\ell^2}\Big], \nonumber \\
C_n(r_+)&=&\frac{dm_n}{dT^n_{\rm
H}}=\frac{\Omega_{n-2}(n-2)r_+^{n-2}}{4G_n}
\Big[\frac{(n-1)r_+^2+(n-3)\ell^2}{(n-1)r_+^2-(n-3)\ell^2}\Big],\label{hbtz2}
 \nonumber \\~S^n_{BH}(r_+)&=&
 \frac{\Omega_{n-2}}{4G_n}r^{n-2}_+,\label{ent2} \\
 T_{\rm H}^n&=&\frac{1}{4\pi r_+}\Big[n-3 +\frac{n-1}{\ell^2}r_+^2\Big],\label{hbtz3} \\
 F_n(r_+)&=& m_n-T^n_{\rm H}
 S_{BH}^n=\frac{\Omega_{n-2}}{16\pi G_n}r_+^{n-3}\Big[1-\frac{r_+^2}{\ell^2}\Big]
  \label{hbtz4}\end{eqnarray}
with the area of $S^{n-2}$ \begin{equation}
\Omega_{n-2}=\frac{2\pi^{\frac{n-1}{2}}}{\Gamma(\frac{n-1}{2})}. \end{equation}
It is easy to check that the first-law of thermodynamics is satisfied as
\begin{equation}\label{dfirst-law} dm_n=T^n_{\rm H} dS^n_{\rm BH}, \end{equation}
where `$d$' denotes the
differentiation with respect to the horizon radius $r_+$ only.
Local thermodynamic (in)stability is determined by the (negative) positive  heat capacity $C_n$ which blows up at $r_+=r^{(n)}_*=\sqrt{\frac{n-3}{n-1}} \ell$.
Fig. 3 indicates a typical shape of the heat capacity  in $n=5$ dimensions.
It turns out that  small AdS black holes with $r_+<r^{(n)}_*$ are  thermodynamically unstable because of $C_n<0$,
while  large AdS black holes with $r_+>r^{(n)}_*$ are  thermodynamically stable because of $C_n>0$.
Classifying an AdS black hole into either  small or large black hole is determined solely by the $n$-dimensional Einstein gravity with a cosmological constant.
\begin{figure*}[t!]
  \centering
 \includegraphics{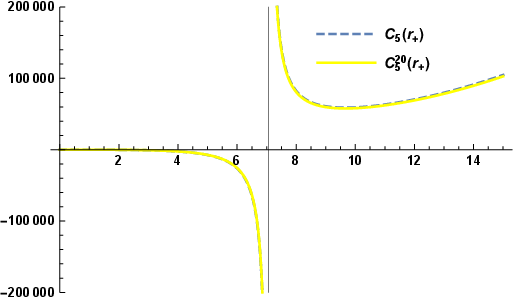}
\caption{Plot for heat capacity $C_{n=5}(r_+)$  with $l=10$ in Einstein gravity and $C^{\alpha=20}_{n=5}(r_+)=0.98C_{n=5}(r_+)$ in Ricci cubic gravity.  Two blow up at $r_+=r_*^{(5)}=\ell/\sqrt{2}=7.1$ (red line).
 The thermodynamic stability is based on the sign of  heat capacity.  The small black hole with $r_+<r_*^{(5)}$ has  the negative heat capacity,
whereas the large black hole with $r_+>r_*^{(5)}$ has the positive heat capacity. This picture persists to $C^{\alpha=20}_{n=5}(r_+)$
in the Ricci cubic gravity.}
\end{figure*}

 Now,  let us  compute the Wald entropy of the AdS black holes to obtain all thermodynamic quantities  in Ricci cubic  gravity.
 The Wald entropy is defined by the following integral performed on $(n-2)$-dimensional spacelike bifurcation surface $\Sigma_{n-2}$~\cite{Ghodsi:2017iee,Wald:1993nt,Iyer:1994ys,Jacobson:1993vj}:
 \begin{eqnarray} \label{w-ent}
 S_{\rm W}&=&-\frac{1}{8} \int_{x_+} d^{n-2}x_+\sqrt{g(r_+)}g^\perp_{\mu\rho}g^\perp_{\nu\sigma} \Big[\frac{\partial {\cal L}_{\rm RC}}{\partial R_{\mu\nu\rho\sigma}}-\bar{\nabla}_\gamma \frac{\partial {\cal L}_{\rm RC}}{\partial \bar{\nabla}_\gamma R_{\mu\nu\rho\sigma}}\Big]^{(0)} \nonumber \\
 &=&-\frac{1}{8} \int_{x_+} d^{n-2}x_+\sqrt{g(r_+)}g^\perp_{\mu\rho}g^\perp_{\nu\sigma}\bar{P}^{\mu\nu\rho\sigma}, \label{w-ent}
 \end{eqnarray}
 where $g^\perp_{\mu\rho}$ represents the metric projection onto the subspace orthogonal to the horizon.
The superscript (0) denotes that the functional derivative with respect to $R_{\mu\nu\rho\sigma}$ is evaluated  on-shell.
 The background (on-shell) $P$-tensor $\bar{P}^{\mu\nu\rho\sigma}$ is given by (\ref{p-tensor}).
Then, the Wald entropy takes the form
\begin{eqnarray}  \label{entropy1}
S_{\rm W}=\frac{A_n}{4}\Big[1+(\alpha+n\beta)a^2\Big]
\end{eqnarray}
with the area of horizon $ A_n=\Omega_{n-2}r_+^{n-2}$ and $\kappa_n=1/G_n=1$.
In  case of $\alpha=-4(n-1)\beta/n$ Ricci cubic gravity, the Wald entropy is given by
\begin{eqnarray}
S_{\rm W}^{\alpha=-4(n-1)\beta/n}= \frac{A_n}{4}\Big[1-\frac{(n-1)(n-2)^2\alpha}{4\ell^4}\Big]. \label{entropy2}
\end{eqnarray}
Up to now, we know the Wald entropy $S^{\alpha=-4(n-1)\beta/n}_{\rm W}$ and  the Hawking temperature $T^n_{\rm H}$ in (\ref{hbtz4}) only.
All thermodynamic  quantities of  $k=0$ AdS  black hole are computed in Ref.\cite{Ghodsi:2017iee}.
In this case, it is known that  there was  a correction $\tilde{\sigma}$ to thermodynamic quantities from Ricci polynomial.
Hence,  we propose that  the first-law  is  satisfied  as
\begin{equation} \label{1st-law}
dM^{\alpha=-4(n-1)\beta/n}=T^n_{\rm H}dS_{\rm W}^{\alpha=-4(n-1)\beta/n }
\end{equation}
in  $\alpha=-4(n-1)\beta/n$ Ricci cubic gravity.
Making use of the first-law (\ref{1st-law}) together with the entropy (\ref{entropy2}) and the Hawking temperature  (\ref{hbtz4}),
one easily  derive the mass as
\begin{eqnarray}
M^{\alpha=-4(n-1)\beta/n }&=&\int^{r_+}_0 T^n_{\rm H}(r_+')dS_{\rm W}^{\alpha=-4(n-1)\beta/n}(r_+') \nonumber \\
&=&\Big[1-\frac{(n-1)(n-2)^2\alpha}{4\ell^4}\Big]m_n(r_+)=\frac{(n-1)\alpha M^2_n}{\ell^2} m_n(r_+), \label{mass}
\end{eqnarray}
where  $M_n^2$ is  the mass squared of massive spin-2 mode in (\ref{MLE-1}).
The other thermodynamic quantities of heat capacity and  free energy  are computed as
\begin{eqnarray} \label{entropy3}
C^{\alpha=-4(n-1)\beta/n }&=&\Big[\frac{dM^{\alpha=-4(n-1)\beta/n }}{dT^n_{\rm H}}\Big]=\frac{(n-1)\alpha M^2_n}{\ell^2}C_n, \label{heat-c} \\
F^{\alpha=-4(n-1)\beta/n}&=&M^{\alpha=-4(n-1)\beta/n }-T^n_{\rm H}S_{\rm W}^{\alpha=-4(n-1)\beta/n}=\frac{(n-1)\alpha M^2_n}{\ell^2}F_n. \label{free-e}
\end{eqnarray}
Now,  we are in a position to mention the thermodynamic (in)stability of the AdS black hole in $\alpha=-4(n-1)\beta/n$ Ricci cubic gravity.
Since the heat capacity $C^{\alpha=-4(n-1)\beta/n}$ blows up at $r_+=r_*^{(n)}$ [see Fig. 3], we   divide still an AdS  black hole into small black hole with $r_+<r_*^{(n)}$ and  large black hole with  $r_+>r_*^{(n)}$. Then, for $M_n^2>0$, it is clear that  small black hole is thermodynamically unstable because  $C^{\alpha=-4(n-1)\beta/n}<0$, while the large black hole is thermodynamically stable  because  $C^{\alpha=-4(n-1)\beta/n}>0$.  For other case of $M_n^2<0$, the situation reverses.  In this case, the small black hole  is thermodynamically stable because
$C^{\alpha=-4(n-1)\beta/n}>0$, while the large black hole is thermodynamically unstable  because  $C^{\alpha=-4(n-1)\beta/n}<0$. However, this case is unacceptable because it corresponds to the unconventional thermodynamic stability.

On the other hand, we find from (\ref{mass-b}) that for $M^2_n>0$,   small (large) black hole with $r_+<r_*^{(n)}(r_+>r_*^{(n)})$ is unstable (stable)  against the $s$-mode massive spin-2 perturbation
 $\delta \hat{R}_{\mu\nu}$. This implies that  the GL instability condition picks up  small AdS black hole which is thermodynamically unstable in $\alpha=-4(n-1)\beta/n$ Ricci cubic gravity.
It shows clearly that the CSC~\cite{Gubser:2000ec} holds for AdS black holes obtained from $\alpha=-4(n-1)\beta/n$ Ricci cubic gravity. However, it is worthwhile to note that the CSC does not hold for   $M_n^2<0$
because it corresponds to the tachyonic instability as well as  its thermodynamic stability is unconventional.

\section{Non-AdS black hole solution}

Before we proceed, we remind the reader that  a static eigenfunction of Lichnerowicz
operator has two roles in Ricci quadratic gravity~\cite{Lu:2017kzi}.  It plays a role of  perturbation away
from Schwarzschild  black hole along a newly non-Schwarzschild black hole, while it plays a role of threshold unstable mode lying at the
edge of a domain of GR instability.  Inspired by this work,
we wish to obtain  a newly non-AdS black hole solution in $\alpha=-4(n-1)\beta/n$ Ricci cubic gravity and determine the threshold mass $M_t$ for GL instability again.
It is worthwhile to note that  this solution is not a full solution to Eq.(\ref{equa1}).

Considering  Eq.(\ref{ansatz}),  we  choose a radial perturbation to obtain  a newly static non-AdS solution
\begin{equation} \label{ansatz1}
ds^2_{\rm pert}=-\bar{f}(r)[1+\epsilon \tilde{h}(r)]dt^2+\frac{dr^2}{\bar{f}(r)[1+\epsilon \tilde{f}(r)]}+r^2d\Omega^2_{n-2},
\end{equation}
where  perturbed metric functions of $\tilde{h}(r)$ and $\tilde{f}(r)$ will be determined by solving linearized equations.
From Eqs.(\ref{MLE-0}) and (\ref{LE-5}), it is known  that the metric perturbation $h_{\mu\nu}={\rm diag}[\bar{f}(r)\tilde{h}(r),\bar{f}(r)\tilde{f}(r),r^2,r^2\sin^2\theta] $ is determined by solving two linearized  equations around AdS black holes
\begin{eqnarray}
&&\delta E^R\equiv\delta R(h_{\mu\nu})=0,\label{pertur1}\\
&&\delta E_{\mu\nu}\equiv(\bar{\Delta}_{\rm L} -2a+M^2_n)\delta\hat{R}_{\mu\nu}(h_{\mu\nu})=0.\label{pertur2}
\end{eqnarray}
Here,  we wish to point out that Eqs.(\ref{pertur1}) and (\ref{pertur2}) are considered as static Lichnerowicz equations in compared to those [Eqs.(\ref{MLE-0}) and (\ref{LE-5})] for GL instability analysis in section 4.
Substituting (\ref{ansatz1}) into Eq.(\ref{pertur1}), one obtains a second-order
coupled equation for $\tilde{h}(r)$ with $\tilde{f}(r)$ as
\begin{eqnarray} \label{pe1}
&&\delta E^R=\Big[-\frac{n(n-1)}{\ell^2}r^2-(n-2)(n-3)\Big]\tilde{f}-\frac{1}{2}\Big[(n-3)r+\frac{n(n-1)}{\ell^2}r^3
+(n-1)r\bar{f}\Big]\tilde{f}'\nonumber\\
&&-\frac{1}{2}\Big[3(n-3)r+\frac{3(n-1)}{\ell^2}r^3-(n-5)r\bar{f}\Big]\tilde{h}'
-r^2\bar{f}\tilde{h}''=0.
\end{eqnarray}
Now, plugging  (\ref{ansatz1}) into Eq.(\ref{pertur2}) leads to the fourth-order  equation which is difficult to be solved directly.
Explicitly, $\delta E_{tt}=0$ and $\delta E_{rr}=0$
become fourth-order  equations for $\tilde{h}(r)$
and third-order equation for $\tilde{f}(r)$, while $\delta E_{\theta\theta}=0$
is a third-order  equation for  $\tilde{f}(r)$ and $\tilde{h}(r)$.
To eliminate a fourth-order term  $\tilde{h}^{''''}(r)$, we combine two component equations  as $\delta E_{tt}+\delta E_{rr}=0$,
leading to a third-order equation for $\tilde{f}(r)$ and $\tilde{h}(r)$.
Furthermore, making use of $\delta E_{\theta\theta}=0$, $\delta E^R=0$ and
 $(\delta E^R)'=0$, the third-order  equation  leads   to a second-order equation
\begin{eqnarray} \label{pe2}
&&-2r^2\left((n-1)r^2/\ell^2+(n-3)\right)\tilde{f}-2r^3\bar{f}\tilde{h}'
+\frac{1}{M_{n}^2}\left[\left((n-3)^2(n-2+2(n-1)r^2/\ell^2)\right.\right.\nonumber\\
&&\left.\left.+(n-4)(n-1)^2r^4/\ell^4
-(n-3)(n(n-1)r^2/\ell^2+(n^2-5n+2))\bar{f}\right)\tilde{f}\right.\nonumber\\
&&\left.+\frac{r}{2}\left(5((n-1)r^2/\ell^2+n-3)^2-(6(n-1)^2r^2/\ell^2+2(n-3)(3n-5))\bar{f}
+(n-1)^2\bar{f}^2\right)\tilde{f}'\right.\nonumber\\
&&\left.+\frac{1}{2}\left(2(n-3)(n-1)r^3\bar{f}/\ell^2
-r((n-1)r^2/\ell^2+n-3)^2+(n-3)\left(2(n-1)\right.\right.\right.\nonumber\\
&&\left.\left.\left.-(n+1)\bar{f}\right)r\bar{f}\right)\tilde{h}'
+\left((n-1)r^2/\ell^2-(n-1)\bar{f}+n-3\right)\tilde{f}''\right]=0.
\end{eqnarray}
Eqs.(\ref{pe1}) and (\ref{pe2}) are solved numerically to find a new static solution when imposing  appropriate boundary conditions.
We note  that the AdS black hole solution $\bar{f}(r)$ approaches asymptotically AdS like  $r^2$ as $r\rightarrow\infty$.
To plug in this asymptotic behavior, we introduce a new  coordinate of  $z=\frac{r_+}{r}$ so that $f(r)$ and $h(r)$
become $f=f(z)$ and $h=h(z)$ whose working  region is compactified  by $0< z \le 1$. Here,   $z=1$ corresponds to  the event horizon $r=r_+$,
whereas $z\to0$ implies  $r\to\infty$.  Hence, we  impose the AdS boundary condition at $z=0$.
However, $f(z)$ and $h(z)$  being proportional to $z^{-2}$ are still divergent at $z=0$.
To resolve it,  functions of $z^2f(z)$ and $z^2h(z)$ are introduced as
\begin{eqnarray} \label{ansatz2}
z^2h(z)=z^2\bar{f}(z)[1+\epsilon \tilde{h}(z)],\quad
z^2f(z)=z^2\bar{f}(z)[1+\epsilon \tilde{f}(z)],
\end{eqnarray}
 which are  finite  between $z=0$ and $z=1$.

Now, we are in a position to rewrite two relevant  equations (\ref{pe1}) and (\ref{pe2}) in terms of $z$ as
\begin{eqnarray} \label{penew1}
&&\Big[\frac{n(n-1)}{\ell^2}\frac{r_+^2}{z^2}+(n-2)(n-3)\Big]\tilde{f}-\frac{z}{2}\Big[(n-3)
+\frac{n(n-1)}{\ell^2}\frac{r_+^2}{z^2}+(n-1)\bar{f}\Big]\tilde{f}'\nonumber\\
&&-\frac{z}{2}\Big[3(n-3)+\frac{3(n-1)}{\ell^2}\frac{r_+^2}{z^2}-(n-9)\bar{f}\Big]\tilde{h}'
+z^2\bar{f}\tilde{h}''=0,
\end{eqnarray}
\begin{eqnarray} \label{penew2}
&&-2\frac{r_+^2}{z^2}\left((n-1)\frac{r_+^2}{z^2\ell^2}+(n-3)\right)\tilde{f}+\frac{2r_+^2}{z}\bar{f}\tilde{h}'
+\frac{1}{M_{n}^2}\left[\left((n-3)^2(n-2+2(n-1)\frac{r_+^2}{z^2\ell^2})\right.\right.\nonumber\\
&&\left.\left.+(n-4)(n-1)^2\frac{r_+^4}{z^4\ell^4}
-(n-3)(n(n-1)\frac{r_+^2}{z^2\ell^2}+(n^2-5n+2))\bar{f}\right)\tilde{f}\right.\nonumber\\
&&\left.-\frac{z}{2}\left(5((n-1)\frac{r_+^2}{z^2\ell^2}+n-3)^2-(\frac{6(n-1)^2r_+^2}{z^2\ell^2}
+2(n-3)(3n-5)+\frac{4(n-1)z^2}{r_+^2})\bar{f}\right.\right.\nonumber\\
&&\left.+(n-1)^2\bar{f}^2-\frac{4(n-1)}{\ell^2}-\frac{4(n-3)z^2}{r_+^2}\right)\tilde{f}'\nonumber\\
&&-\frac{z}{2}\left(2(n-3)(n-1)\frac{r_+^2}{z^2\ell^2}\bar{f}
-((n-1)\frac{r_+^2}{z^2\ell^2}+n-3)^2+(n-3)\left(2(n-1)\right.\right.\nonumber\\
&&\left.\left.\left.-(n+1)\bar{f}\right)\bar{f} \right)\tilde{h}'
+\left(\frac{(n-1)z^2}{\ell^2}-\frac{(n-1)z^4}{r_+^2}\bar{f}+\frac{(n-3)z^4}{r_+^2}\right)\tilde{f}''\right]=0.
\end{eqnarray}
Here, the prime ($'$) denotes a derivative with respect to $z$.
\begin{figure*}[t!]
   \centering
   \includegraphics{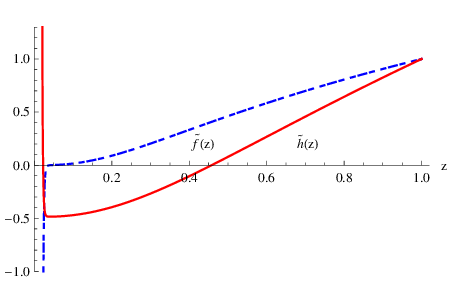}
   \hfill%
   \includegraphics{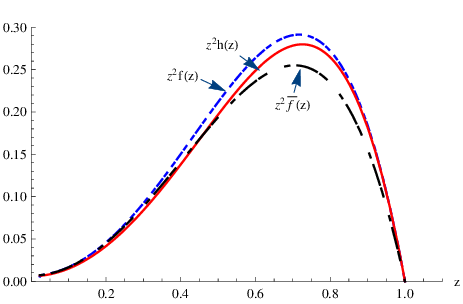}
\caption{Two graphs for a newly non-AdS black hole solution  with $\ell=10$, $r_+=1$,  and $\epsilon=0.1$ in $n=5$ dimensions.
The left picture shows $\tilde{f}(z)$ and $\tilde{h}(z)$ as function of $z\in (0,1]$ for a non-AdS black hole. The  AdS infinity appears  at $z=0$,
while the event horizon is located at $z=1$.  The right picture indicates clearly  a distinction between AdS
and non-AdS black holes: $z^2f(z)=z^2h(z)=z^2\bar{f}(z)$ for the former and $z^2f(z)\not=z^2h(z)$ for the latter.}
\end{figure*}

In order to obtain a regular solution on the horizon $z=1$, it is standard  to introduce $\tilde{f}(z)$ and $\tilde{h}(z)$  as the Taylor expansion
\begin{eqnarray} \label{expansion1}
\tilde{f}(z)&=&1+f_1(1-z)+f_2(1-z)^2+\cdots, \\
\tilde{h}(z)&=&1+h_1(1-z)+h_2(1-z)^2+\cdots. \label{expansion2}
\end{eqnarray}
Substituting (\ref{expansion1}) and  (\ref{expansion2}) into (\ref{penew1}) and (\ref{penew2}),  two relevant  coefficients of  $f_1$ and $h_1$ are determined  to be
\begin{eqnarray} \label{coeff}
&&f_1=\frac{3M_n^2r_+^2-2(n-3)\Big[(n-1)r_+^2/\ell^2+(n-2)\Big]}{4\Big[(n-1)r_+^2/\ell^2+(n-3)\Big]},\nonumber\\
&&h_1=-\frac{M_n^2r_+^2+2\Big[(n^2-1)r_+^2/\ell^2+(n-2)(n-3)\Big]}{4\Big[(n-1)r_+^2/\ell^2+(n-3)\Big]}.
\end{eqnarray}
We emphasize that for a given $r_+$, a newly static non-AdS solution is allowed  only  for  a particular value of $\tilde{M}_n$.
For $r_+=1$ and $\ell=10$, we obtain $\tilde{M}_n$ depending on $n$ as
\begin{equation} \label{qwn}
\left[
\begin{array}{c|ccccccc}
 n & 4 & 5 & 6 & 7 & 8 & 9 & 10 \cr \hline
      \tilde{M}_n & 0.86 & 1.26 & 1.57 & 1.83 & 2.07 & 2.29& 2.49 \cr
       \tilde{M}_n^{2} & 0.74 & 1.59 & 2.46 & 3.45 & 4.28 & 5.24& 6.2 \cr
\end{array}
\right].
\end{equation}
We note that $\tilde{M}_n$ is exactly the threshold mass  $M^t_n$ found  in  Eq.(\ref{cwn}) for  GL instability.
It describes  a close connection between the GL instability of AdS black holes and  newly non-AdS black holes.
This is so because we have solved the same linearized equations for  static metric perturbations with $\Omega=0$.
In Fig. 4, we depict  a newly non-AdS  black hole solution in $n=5$ dimensions by comparing with AdS black hole.
It corresponds to a threshold unstable mode lying at the
edge of a domain of GR instability for  a small AdS black hole.
At this stage, we would like to mention that  a non-SAdS black hole in $n=4$ dimensions was
found in Ricci quadratic gravity with a cosmological constant~\cite{Lin:2016kip}.

\section{Discussions}

In the AdS black hole background of quasi-topological gravity (\ref{QTG}), the AdS black hole is stable against the metric perturbation $h_{\mu\nu}$ because its linearized quasi-topological gravity is given by $\delta G_{\mu\nu}=0$.

 We note that the linearized $\alpha=-4(n-1)\beta/n$ Ricci cubic gravity  is described by  the Lichnerowicz equation (\ref{LE-5}).  It includes a massive spin-2 mode ($\delta \hat{R}_{\mu\nu}$) and thus,  small ($r_+<r_*^{(n)}$) and large ($r_+>r_*^{(n)}$ ) AdS black holes are unstable and stable against the Ricci tensor perturbation $\delta \hat{R}_{\mu\nu}=e^{\Omega t}\{\cdots\}_{\mu\nu}$.
The former corresponds to the GL instability.
On the other hand, a small AdS  black hole ($r_+<r_*^{(n)}$) is thermodynamically unstable in the canonical ensemble because its heat capacity is negative, whereas  a large AdS  black hole ($r_+>r_*^{(n)}$) is thermodynamically stable  because its heat capacity is positive.  This implies that  the correlated stability conjecture (CSC) holds for the AdS black holes  in $\alpha=-4(n-1)\beta/n$  Ricci cubic gravity.
But the CSC does not hold for AdS black holes in quasi-topological  gravity.

Furthermore, we have obtained  a newly non-AdS black hole solution in the $\alpha=-4(n-1)\beta/n$ Ricci cubic  gravity numerically
by solving   static  Lichnerowicz equations.  The solution is allowed only for $\tilde{M}_n=M_n^t$, which confirms the threshold mass $M^t_n$  for the GL instability.

Recently, it was shown that the non-Schwarzschild black hole is unstable against the metric perturbations in the second-order formalism of Ricci quadratic gravity when the mass bound of spin-2 mode is
satisfied~\cite{Held:2022abx}
 \begin{equation}
 0<m_2<\frac{0.87}{r_0},
\end{equation}
which is the same bound of   GL instability for Schwarzschild(-AdS) black hole as in Eq.(\ref{kwn}). 
However, this result implies that the non-Schwarzschild black hole (a black hole with Ricci-tensor hair) might  not be survived as a newly physical black hole because it is unstable in the same branch of GL instability for Schwarzschild black hole.  On the other hand, there was a significant progress on obtaining black holes with scalar hair via spontaneous scalarization.
In this case,  the tachyonic instability of  black holes is regarded as  the hallmark for emerging  scalarized black holes when introducing scalar coupling $f(\phi)$  to the source term: the Gauss-Bonnet term ($R^2-4R_{\mu\nu}R^{\mu\nu}+R_{\mu\nu\rho\sigma}R^{\mu\nu\rho\sigma}$)   for Schwarzschild black hole~\cite{Doneva:2017bvd,Silva:2017uqg,Antoniou:2017acq}  or Maxwell term ($F^2$) for Reissner-Nordstr\"{o}m black hole~\cite{Herdeiro:2018wub}. Infinite branches of $n=0,~1,~2,\cdots$ scalarized black holes were found  in the unstable branch of GR black holes. The  fundamental ($n=0$) branch is stable, whereas all excited ($n=1,~2,\cdots$) blanches are unstable~\cite{Zou:2020zxq}. This suggests  that the fundamental branch is considered as the end point of GR black holes through the Hawking radiation.

 \vspace{1cm}

{\bf Acknowledgments}
 
\end{document}